\documentclass[prl,amssymb,twocolumn,superscriptaddress,showpacs,floatfix]{revtex4}

\usepackage{graphicx}

\begin{document}

\title{Self-tuned quantum dot gain in photonic crystal lasers}

\author{S. Strauf} \email[Corresponding author: ]{strauf@physics.ucsb.edu}
\affiliation{Materials Department, University of California Santa Barbara, CA 93106, Santa Barbara, USA}
\affiliation{Department of Physics, University of California Santa Barbara, CA 93106, Santa Barbara, USA}

\author{K. Hennessy}
\affiliation{ECE Department, University of California Santa Barbara, CA 93106, Santa Barbara, USA}

\author{M. T. Rakher}
\affiliation{Department of Physics, University of California Santa Barbara, CA 93106, Santa Barbara, USA}

\author{Y.-S. Choi}
\affiliation{ECE Department, University of California Santa Barbara, CA 93106, Santa Barbara, USA}

\author{A. Badolato}
\affiliation{ECE Department, University of California Santa Barbara, CA 93106, Santa Barbara, USA}

\author{L. C. Andreani}
\affiliation{Department of Physics ``A. Volta'', University of Pavia, 27100 Pavia, Italy}

\author{E. L. Hu}
\affiliation{Materials Department, University of California Santa Barbara, CA 93106, Santa Barbara, USA}
\affiliation{ECE Department, University of California Santa Barbara, CA 93106, Santa Barbara, USA}

\author{P. M. Petroff}
\affiliation{Materials Department, University of California Santa Barbara, CA 93106, Santa Barbara, USA}
\affiliation{ECE Department, University of California Santa Barbara, CA 93106, Santa Barbara, USA}

\author{D. Bouwmeester}
\affiliation{Department of Physics, University of California Santa Barbara, CA 93106, Santa Barbara, USA}


\begin{abstract}
We demonstrate that very few (1 to 3) quantum dots as a gain medium are sufficient to realize a photonic crystal laser
based on a high-quality nanocavity. Photon correlation measurements show a transition from a thermal to a coherent
light state proving that lasing action occurs at ultra-low thresholds. Observation of lasing is unexpected since the
cavity mode is in general not resonant with the discrete quantum dot states and emission at those frequencies is
suppressed. In this situation, the quasi-continuous quantum dot states become crucial since they provide an
energy-transfer channel into the lasing mode, effectively leading to a {\it self-tuned} resonance for the gain medium.
\end{abstract}

\maketitle

Optical microcavities \cite{Vahala:Nature2003}, in particular photonic-crystal membranes, offer the ability to create
new, efficient optical sources of specified wavelength through the control of the dielectric environment. Extremely
low-threshold lasers can result from the appropriate match of small mode-volume photonic-crystal cavities to optically
active material. One measure of the optical efficiency of a laser is the spontaneous emission (SE) coupling factor
$\beta$. The theoretical limit is one, corresponding to the hypothetical case of thresholdless lasing. Microdisk
\cite{Slusher:APL93,Zhang:APL03} and photonic-crystal lasers (PCLs) \cite{Ryu:APL04} with quantum wells (QWs) or a high
density of quantum dots (QD) as gain medium show values ranging from 0.1 to 0.2, but a pronounced lasing threshold
behavior remains \cite{Slusher:APL93,Zhang:APL03,Ryu:APL04,Painter:Science99,Michler:APL00}. For non-lasing devices
$\beta$ values above 0.9 have been estimated \cite{Kress:PRB05}. Here we report on PCLs
sustained by only 1 to 3 QDs as active gain material showing ultra-low lasing thresholds and a $\beta$ of 0.85.\\
\indent To achieve lasing operation, both the gain medium and the cavity must have exceptional features. The cavity
design shown in Fig.~1a is based on a line defect of three missing air holes (L3-type) within a triangular lattice
formed in a 126 nm GaAs membrane, acting as a mono-mode waveguide in the QD emission range. The center of the membrane
contains a single layer of low-density (5-50 ${\mu \rm m}^{-2}$) InAs QDs grown by the partial covered island technique
\cite{Garcia:APL98}. This leads to QDs with shallow confinement energies and a pronounced density of extended wetting
layer (WL) states \cite{Vasanelli:PRL2002}.
\begin{figure}[tb!]
\includegraphics[width=83mm]{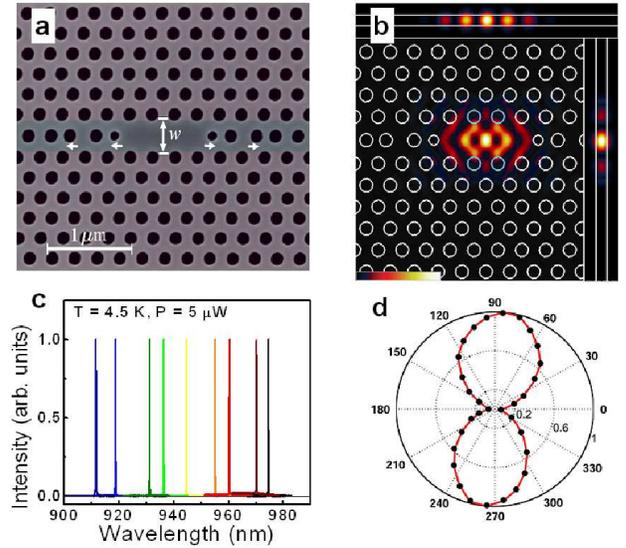}
\caption{(a) Scanning electron micrograph around the defect region of a PCL laser with a lattice constant of 260 nm.
(b) Plot of the electric field intensity of the lasing mode as calculated from 3D finite-difference time-domain
simulations \cite{KevinAPL:05}. (c) Normalized lasing mode spectra for 9 different PCL devices taken at a pump power of
5 $\mu$W, T = 4.5 K. (d) Measured cavity mode intensity as a function of polarizer angle showing a polarization ratio
of 25:1.}
\end{figure}
Neighboring holes in the line have been shifted outwards (as indicated by white arrows in Fig.~1a) to produce better
confinement of the mode \cite{Akahane:Nature03}. To achieve optimal coupling of the lasing mode with the active QD gain
material, two new design concepts have been employed. First, the two holes nearest to the defect have been reduced by
20\% in size to decrease their overlap with high field regions \cite{Andreani:PNFA04}. Second, the width $w$ of the
channel region containing the three missing holes has been increased by 30 nm, as illustrated by the blue shading in
Fig.~1a. In culmination, these design concepts achieve a mode volume of $V = 0.68~(\lambda/n)^3$ and an effective
refractive index $n_{eff}$ of 2.9, a very high value for this class of porous cavity membranes. In addition, this
design ensures that the field maximum is in excellent spatial overlap with the gain medium while the overlap with the
air-semiconductor interface is drastically reduced. Such interfaces are detrimental since the optical properties of QDs
near surfaces
degrade critically \cite{Wang:APL04}, and surface defect states can be optically absorbent.\\
\indent The wavelength of the PCL mode can be tuned through its lattice parameters to the SE wavelength range of the QD
gain medium. We studied 22 PCL devices ranging in wavelength from 908 to 975 nm covering the ground state (s-shell,
centered at 965 nm) and excited state (p-shell) of the embedded QDs. The QDs were excited non-resonantly via the GaAs
matrix by a continuous wave (cw) 780 nm laser diode focused to a 2 $\mu m$ spot size using a microscope objective.
Despite the fact that there are on average only 1 to 3 QDs located within the mode volume and that the exciton
transitions of those QDs are in general off resonance with the cavity mode, we claim that most devices show
pronounced single mode lasing (Fig.~1c-d).\\
\indent Before discussing the remarkable laser gain process, we present three-fold experimental evidence for lasing
operation of these devices. First, we recorded the output power of the cavity mode as a function of cw-excitation pump
power. Figure~2a shows the resulting L-L-curve for a device emitting at 960 nm (solid dots).
\begin{figure}[tb!]
\includegraphics[width=85mm]{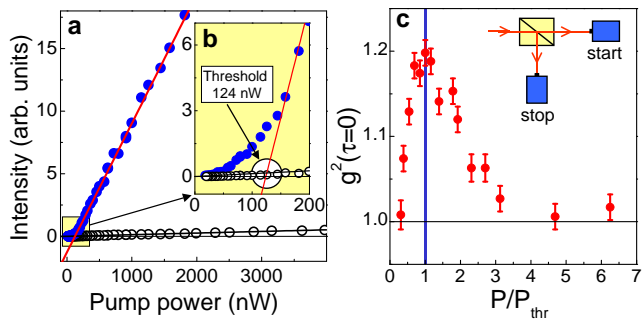}
\caption{(a) Lasing mode intensity (solid dots) and SE background (open dots) as a function of excitation pump power
for a cavity emitting at 960 nm. Red line: linear fit. (b) Magnification of threshold region. (c) Photon correlation
function $g^2(\tau=0)$ of the cavity mode as a function of normalized pump power $P/P_{thr}$. Absorbed pump power at
threshold, $P_{thr}$, for this device emitting at 940 nm is 25 nW. T = 4.5 K.}
\end{figure}
Above a certain threshold pump power, a linear increase in output intensity is observed, as indicated by the red line.
The conventional estimation of the cw lasing threshold, obtained by extrapolation of the red line to zero output power,
yields a value of 124 nW (Fig.~2b) corresponding to an absorbed power of only 4 nW (14 $\rm mW/cm^2$). This is an
improvement of 2 to 3 orders of magnitude over prior reports
\cite{Vahala:Nature2003,Slusher:APL93,Zhang:APL03,Ryu:APL04,Painter:Science99,Michler:APL00}. We measured similar
ultra-low threshold values for temperatures between 4 and 80 K. For higher temperatures the QD emission is quenched due
to the thermal break up of excitons. Figure 2b shows no sharp threshold but a soft turn-on of the cavity mode
intensity,
in accordance with theory for a large SE coupling efficiency \cite{Bjork:PRA1994,Exter:PRA1996}.\\
\indent A second characteristic of a laser is the emission linewidth narrowing profile. Below and above threshold the
linewidth $\Delta E$ varies inversely with the output power \cite{Coldren:Book1995}. At threshold, the phase transition
into lasing leads to a pronounced kink in the profile \cite{Bjork:APL92}, as demonstrated in Fig.~3a.
\begin{figure}[tb!]
\includegraphics[width=63mm]{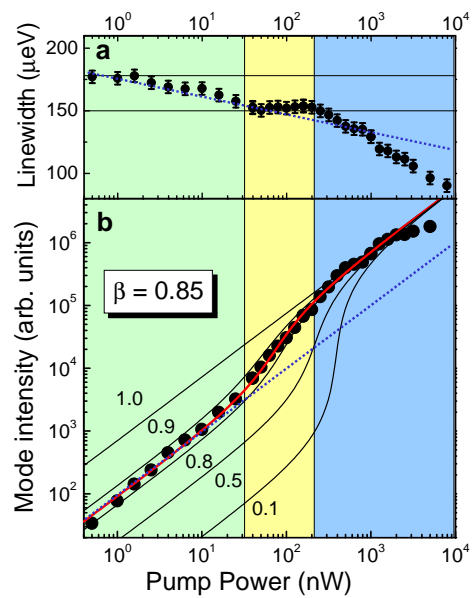}
\caption{Cavity mode linewidth (a) and intensity (b) as a function of excitation pump power (solid dots). The x-axis
displays optical power sent into the cavity. The corresponding absorbed pump powers are ~3.6 \% of these values. Solid
lines: Rate equation model solutions for various values of $\beta$. Red line: Best fit corresponding to $\beta =
0.85\pm0.03$. Dotted blue lines indicate the measured behavior for non-lasing devices.}
\end{figure}
To compare the linewidth narrowing with the intensity data (Fig.~3b) the linewidth is plotted as a function of pump
power. In principal, a contribution to the observed linewidth narrowing could also arise from loss saturation effects
of QDs. These effects, however, do not lead to a kink in the L-L curve, as has been confirmed by additional
measurements performed on non-lasing devices. Those measurements follow the dotted blue lines in Fig.~3. Such behavior
corresponds to operation in the light-emitting-diode (LED) regime and is similar to the standard Schawlow-Townes
linewidth narrowing below threshold, where there is a constant cavity loss but an increase in gain
\cite{Coldren:Book1995}. At pump powers of 3-5 $\mu$W, an overall saturation of the cavity mode intensity sets in due
to saturation of the QD gain medium. We found that maximum output powers for devices operating in the LED regime are
typically an order of magnitude lower compared to those reaching into the lasing regime. For the best lasing devices
the $E/ \Delta E$ ratios
reach values up to 32000 and the corresponding Q factors of the cold cavity range between 7800 and 18000.\\
\indent To provide a third and direct proof that lasing action occurs, we measured the second order photon correlation
function $g^2(\tau)={\langle I(t) I(t+\tau) \rangle / \langle I(t) \rangle  \langle I(t) \rangle}$, where $\langle I(t)
\rangle$ is the expectation value of the intensity in the cavity mode at time t, as a function of excitation pump power
(Fig.~2c). The cavity mode emission was spectrally filtered with a 0.5 nm bandpass and sent into a Hanbury-Brown and
Twiss (HBT) interferometer, consisting of a 50/50 beamsplitter and two single photon counting detectors, as shown
schematically in the inset of Fig.~2c. Photon coincidences were recorded electronically as a histogram of start-stop
events. A characteristic feature of a laser is the phase transition from a thermal into a coherent light state as a
function of pump rate \cite{Jin:PRA1994}. The measured photon bunching signature $(g^2(\tau=0)>g^2(\tau))$ around the
threshold of the PCL indicates the thermal character of the emitted light field. Below threshold $g^2(\tau=0)$ should
theoretically approach a value of two corresponding to an ideal thermal state. However, the measured values for
$g^2(\tau=0)$ drop down at lowest pump powers due to a decrease of the mode signal to SE background ratio and due to a
decrease of the coherence length of the PCL approaching the temporal resolution limit (600 ps) of the HBT setup
\cite{Deng:2002}. At pump powers above threshold the cavity mode picks up phase coherence due to stimulated emission,
leading to a transition into a coherent light state where $g^2(\tau=0)$ approaches unity (Fig.~2c). In addition, we
performed experiments on a series of photonic crystal cavities designed to span the range from $\beta=0.3$ to
$\beta=0.9$ as well as on a commercial semiconductor laser diode confirming that Fig.~2c is indeed
characteristic for lasing action \cite{Yong}.\\
\indent To extract the fraction $\beta$ of SE into the lasing mode with respect to SE into all available modes, we use
a coupled rate equation model for the carrier density $N$ and the photon density $P$ of a semiconductor laser
diode\cite{Coldren:Book1995}:
\begin{eqnarray}
\frac{dN}{dt} &=& R_p - \frac{N}{\tau_{sp}} - \frac{N}{\tau_{nr}} - v_g g P\\
\frac{dP}{dt} &=& v_g \Gamma g P + \Gamma \beta \frac{N}{\tau_{sp}}- \frac{P}{t_c}
\end{eqnarray}
where $R_p$ is the pump rate, $v_g$ the group velocity, $\Gamma$ the confinement factor, $1/t_c$ the cavity photon loss
rate, $1/t_{sp}$ the spontaneous, and $1/t_{nr}$ the non-radiative recombination rate. Steady state solutions have been
found assuming a linear gain function $g=a(N-N_{tr})$, where $a$ is the differential gain and $N_{tr}$ the transparency
carrier density. The solid lines shown in Fig.~3b are solutions of the coupled rate equation model for various values
of $\beta$ between 0.1 and 1.0 keeping all other parameters constant. The parameters used are determined as follows: A
$t_c$ of 4.1 ps corresponds to the measured Q = 9800. Lifetime measurements at the cavity mode frequency reveal
$t_{sp}=0.1$~ns. The $t_{nr}$ has been taken to be 10 ns and does not influence the fit considerably
\cite{Bjork:PRA1994}. A $\Gamma$ of 0.004 has been estimated from the intersection of the mode volume with the single
layer of QDs and $v_g=c/n_{eff}=1\times 10^{10}$ cm/s. Best fits have been found for $a = 4.5 \times 10^{-10} \rm cm^2$
and $N_{tr} = 5.5 \times 10^{14} \rm cm^{-3}$. The corresponding material gain of $2\times 10^{5} \rm cm^{-1}$ is in
good agreement with values for other InAs QD lasers \cite{Bimberg:1997}. The transparency (threshold) carrier density
corresponds to about 8 (18) electron hole pairs within the mode volume. The red curve in Fig.~3b is the best fit for
$\beta = 0.85\pm0.03$ which implies an ultra-high SE coupling efficiency of 85\%. Note that the nonlinear kink in the
intensity data occurs at the same pump power values as the kink in the linewidth narrowing (Fig.~3a) and the soft
turn-on starting at about 30 nW incident (1 nW absorbed) pump power corresponds to the definition of the lasing
threshold for high-$\beta$ lasers
\cite{Bjork:PRA1994}.\\
\indent In general, the $\beta$ factor depends on the number of optical modes and the width of the SE spectrum. Besides
the fundamental cavity mode, the density of other radiation modes is strongly suppressed. This is supported by measured
lifetimes up to 10 ns for single QDs located spatially and spectrally outside of the cavity mode region but within the
photonic crystal membrane. These values are one order of magnitude larger than typical SE lifetimes (1~ns) measured for
single QDs in the unprocessed parts of the wafer. The observed inhibition of SE directly reflects the low density of
radiation modes within the photonic bandgap. It has been predicted that non-degenerate fundamental modes in PCL
structures with mode profiles similar to that of our design can achieve $\beta$ values up to 0.87 assuming a SE
bandwidth of 25 nm \cite{Vuckovic:JQE1999}. While such a bandwidth can be achieved with QWs or larger ensembles of QDs
\cite{Michler:APL00}, it is indeed very surprising that at our low QD densities (1 to 3 QDs spatially within the mode
volume) such an efficient QD-cavity coupling occurs. The sharp exciton QD emission is furthermore statistically
distributed over 50 nm making a simultaneous spectral and spatial coupling of a single QD very unlikely ($<1\%$).
Control samples containing zero QDs but a dominant WL emission at 860 nm do not show lasing or any decoration of the
cavity mode. If there is only one QD spatially overlapping with the cavity mode, as has been confirmed by deterministic
QD positioning \cite{Badolato:Science05}, decoration of the mode occurs even for up 20 nm spectral detuning
\cite{KevinAPL:05}.
Therefore, the quantum dot cavity interaction has to be of an indirect nature.\\
\indent To obtain insight into the underlying gain mechanism, we studied the spectral properties of single QDs in an
unprocessed part of the wafer (Fig.~4a-c).
\begin{figure}[tb!]
\includegraphics[width=60mm]{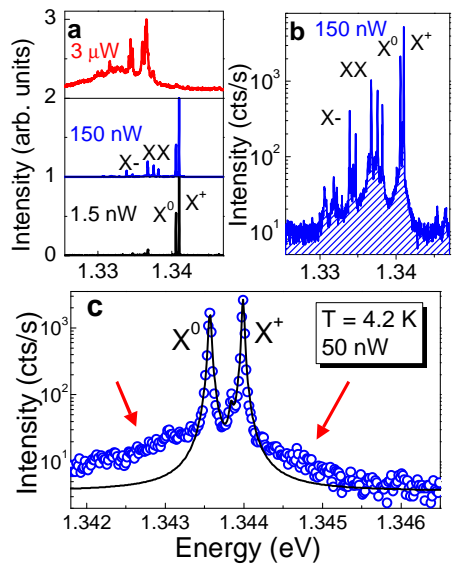}
\caption{(a) Single QD spectra recorded at pump powers below (1.5nW), around (150 nW) and above (1.5 $\mu$W) the lasing
threshold of the PCLs. (b) 150 nW spectrum on a log scale illustrating the broad background. (c) High-resolution
spectrum taken at 50 nW (blue dots). Solid line: Lorentzian fits of the $X^0$ and $X^+$ transitions. The red arrows
highlight deviations caused by acoustic phonon coupling.}
\end{figure}
The sharp transitions of the s-shell emission evolve with increasing pump power into spectrally broad bands (Fig.~4a).
This broad background is significant at pump powers around the lasing threshold (Fig.~4b) and demonstrates that SE from
a single QD occurs in a large frequency range. Several effects are responsible for the rich spectral properties.
Multiple charge states of the exciton ($X^+, X^0, X^-$) \cite{Warburton:Nature2000} and higher order Coulomb
correlations like bi-excitons (XX) \cite{Moreau:PRL2001} give rise to several recombination lines (Fig.~4a). Each
transition deviates from a Lorentzian line shape in the form of a broad background extending about 4 meV (Fig.~4c) due
to electron-acoustic phonon coupling \cite{Besombes:PRB2001}. It is furthermore known that in these shallowly confined
QDs the interaction between localized QD states and extended states of the adjacent WL provide a broad quasi-continuum
spectrally covering the energy region between the WL and the p-shell transitions \cite{Vasanelli:PRL2002}. These states
are already appreciably populated at pump powers around the lasing threshold. In this case, when multi-exciton states
are involved, the recombination energy of the s-shell strongly depends on the exact exciton number and spin
configuration in the p-shell \cite{Bayer:Nature2000} and extended states. Therefore, this Coulomb interaction
effectively maps the extended state continuum around the p-shell into the energy region of the s-shell and contributes
to a broad emission background as shown in Fig~4a top.\\
\indent If the QD exciton is spatially but not spectrally coupled to the cavity mode, the transition for the s-shell
exciton is strongly inhibited ($\tau=10~ns$) and thus almost always populated. In this situation, {\it additional}
excitons are forced to populate the extended state continuum and therefore provide a transfer channel into the lasing
mode through their Coulomb interaction with the s-shell. Carrier relaxation within the extended states of QDs has
recently been measured to be faster than 10~ps \cite{Bogaart:2005}, which is comparable to the photon hold time
$\tau_c$ of the PCL devices. In culmination, the QD confinement potential harvests the excitons and their coupling to
the surrounding matrix (extended WL states and acoustic phonons) efficiently transfers the energy into the lasing mode.\\
\indent In summary, we observe lasing action as demonstrated by the characteristic L-L curve, linewidth narrowing
profile and photon statistics. We have shown that lasing in single-mode photonic-crystal cavities does not require an
exciton/mode resonance. These results challenge the conventional QD-laser design based on incorporating several dense
layers of QD material. This has clear technological implications for future design of ultra-efficient nanoscale lasers,
which may now be formed with only a few, or even single, QDs that, with cooperation from the surrounding matrix, {\it
self-tune} their emission into the lasing mode with nearly perfect efficiency.\\
This research has been supported by DARPA Grant No. MDA972-01-1-0027 and NSF NIRT Grant No. 0304678.


\end{document}